\documentclass[amsmath,amssymb, aps, prx, longbibliography, twocolumn]{revtex4-1}

\usepackage{times}
\usepackage{natbib}
\usepackage{graphicx}
\usepackage{dcolumn}
\usepackage{bm}
\usepackage[usenames]{color}
\usepackage{subfigure}

\begin{document}

\title{ Observation of non-Fermi liquid physics in a quantum critical metal \\ via quantum loop topography}

\author{George (Trey) Driskell$^1$}
\author{Samuel Lederer$^1$}
\author{Carsten Bauer$^2$}
\author{Simon Trebst$^2$}
\author{Eun-Ah Kim$^1$}
\email{eun-ah.kim@cornell.edu}

\affiliation{$^{1}$Department of Physics, Cornell University, Ithaca, New York 14853, USA}
\affiliation{$^{2}$Institute for Theoretical Physics, University of Cologne, 50937 Cologne, Germany}

\date{\today}

\begin{abstract}
Non-Fermi liquid  physics is a ubiquitous feature in strongly correlated metals, manifesting itself in anomalous transport properties, such as a $T$-linear resistivity in experiments. However, its theoretical understanding in terms of microscopic models is lacking despite decades of conceptual work and attempted numerical simulations. 
Here we demonstrate that a combination of sign problem-free quantum Monte Carlo sampling and quantum loop topography,
a physics-inspired machine learning approach, 
can map out the emergence of non-Fermi liquid physics in the vicinity of a quantum critical point with little prior knowledge. 
Using only three parameter points for training the underlying neural network, we are able to reproducibly identify a stable non-Fermi liquid  regime tracing the fan of a metallic quantum critical points at the onset of both spin-density wave and nematic order.
Our study thereby provides an important proof-of-principle example that new physics can be detected via unbiased machine-learning approaches.
\end{abstract}

\maketitle

Correlated electrons can give rise to a wide range of different macroscopic quantum phenomena, yet there is one recurring quantum many-body state that is of central importance -- the formation of a non-Fermi liquid (NFL), which is found in the vicinity of such distinct states as quantum critical metals, superconductors, 
or fractionalized quantum matter. 
Conceptually, NFLs are systems of interacting electrons that evade a description in terms of Landau's Fermi liquid theory of metals~\cite{haldane1994,ong2001more}. Experimentally, this is often established via the observation of deviations
from Fermi liquid phenomenology, such as the absence of a constant specific-heat coefficient or, more strikingly, in transport measurements that show a deviation from a $T^2$ dependence of the resistivity at low temperatures \cite{lohneysen2007}.
In fact, the observation of an almost perfect $T$-linear resistivity above the superconducting dome (masking a quantum critical point) 
in a number of $3d$ 
transition metal oxides has been an experimental hallmark of what is widely dubbed a ``strange metal" regime 
\cite{SachdevKeimer2011}. 
Developing a
theoretical understanding of such non-Fermi liquid physics and establishing its microscopic origin 
has, however, remained one of the major outstanding challenges of condensed matter theory over the past few decades.

Conceptual difficulties in studying NFL physics near quantum critical points have arisen along multiple fronts:
(i) The interaction of gapless modes of the bosonic order parameter with the profusion of Fermi surface excitations has
hampered efforts towards a controlled analytical treatment \cite{PhysRevLett.82.4280, PhysRevLett.98.136402, PhysRevLett.106.106403, PhysRevB.89.155130, PhysRevB.90.165146, PhysRevB.99.235136, PhysRevB.100.115103, VIEIRA2020168230}.
(ii) The numerical analysis of many-electron systems has seen similar road blocks -- either in the form of the sign problem \cite{Loh1990} in fermionic quantum Monte Carlo (QMC) approaches or the accelerated growth of entanglement \cite{Klich2006,Wolf2006} in tensor network approaches.
(iii) Even in models that permit some exact understanding, the characteristic transport quantities revealing NFL physics are notoriously difficult to calculate. This is because they are intrinsically non-equilibrium properties, and therefore require, for instance, the analytical continuation of imaginary-time correlations to real time in QMC approaches (a numerically ill-posed problem). Moreoever, it is impossible to break down a transport quantity in terms of theoretically accessible Green's functions unless vertex corrections vanish under the model assumptions \cite{parcollet1999,cha2019tlinear,cha2020linear}.  The conundrum is that it is the transport experiments that strikingly anchor the NFL regions to the quantum critical points in most experiments \cite{SachdevKeimer2011}.

Recently, however, there has been significant progress in cracking some of the numerical barriers. In a vanguard line of research, a new
family of microscopic spin-fermion models has been formulated \cite{Berg2012} that, by construction, are devoid of the infamous sign problem \cite{Berg2019, Xu_2019}. These models have allowed for numerically exact studies of quantum criticality in metals undergoing phase transitions to spin-density wave (SDW) \cite{Schattner2016,Gerlach2017,Liu2018}, nematic \cite{PhysRevX.6.031028,Lederer4905}, ferromagnetic \cite{PhysRevX.7.031058, xu2020extracting}, or ``clock" \cite{PhysRevB.98.045116} orders, often accompanied by the concurrent formation of superconductivity.
The possibility of NFL physics in the vicinity of 
such quantum critical behavior has been preliminarily explored in numerical experiments 
using imaginary-time proxy observables for quantities such as the  quasiparticle spectral weight \cite{PhysRevX.6.031028,Schattner2016}, the Fermi velocity \cite{Gerlach2017}, or the DC resistivity \cite{Lederer4905}.
While such proxies seem to indicate
a breakdown of Fermi liquid behavior, more direct measures of transport phenomena are 
desired to conclusively map out the putative NFL regime in the finite-temperature phase diagrams of these models.
As a first step in this direction, it has recently been argued that a combination of QMC sampling and quantum loop topography (QLT) -- a physics-inspired machine learning algorithm -- is capable of quantitatively probing transport properties \cite{Bauer2020}. Proof-of-principle calculations of this QMC+QLT approach, probing the onset of superconductivity in the attractive Hubbard model and one the aforementioned spin-fermion models, has yielded striking consistency with numerically exact results for these systems \cite{Bauer2020}.

 \begin{figure*}[t]
    \centering
	\includegraphics[width=\textwidth]{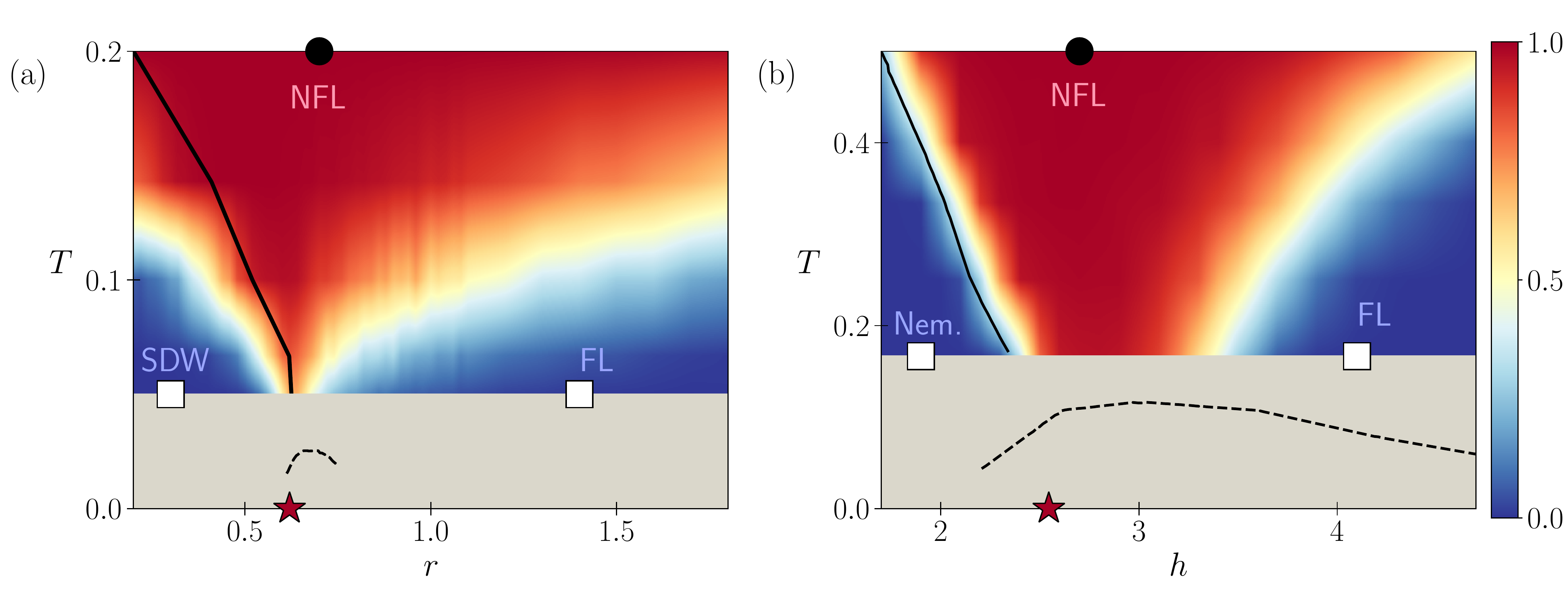}
    \caption{
    {\bf Machine learning of non-Fermi liquid physics.}
    Phase diagrams of quantum critical metals overlaid with machine-learned Fermi liquid to non-Fermi liquid crossover. The color maps show the output of the neural networks trained to classify Fermi liquid and non-Fermi liquid regimes of the spin density wave model (a: Eq. \ref{Eq:SDW}, with $\lambda =1.5,c=3,u=1,\mu=-0.5$) and the nematic model (b: Eq. \ref{Eq:Nematic}, with $\alpha=1.5,V=0.5,\mu=-1$).
    A value of 1 (dark red) corresponds to the non-Fermi liquid, a value of 0 (dark blue) corresponds to the Fermi liquid, and intermediate values represent the crossover region. The neural networks are trained on samples of the equal-time Green's function drawn from quantum Monte Carlo simulations of the models preprocessed via Quantum Loop Topography (QLT) for the SDW model and both QLT and nearest neighbor Green's function data for the nematic model at the training points shown in the figure: white boxes for the Fermi liquid (a: $r=0.3,T=0.05$ and $r=1.4,T=0.05$; b: $h=1.9,T=0.17$ and $h=4.1,T=0.17$) and black circles for the non-Fermi liquid (a: $r=0.7,T=0.2$; b: $h=2.7,T=0.5$). The red stars are placed at quantum critical value of the tuning parameters (a: $r_{c}=0.62$; b: $h_{c}=2.6$). The solid black lines show the phase boundaries, and the dashed black lines show the superconducting $T_{c}$, from refs. \cite{Lederer4905,Gerlach2017}.}
    \label{fig:pds}
\end{figure*}

In this manuscript, we demonstrate that this combination of machine-learning assisted analysis and sign-problem free Monte Carlo sampling (QMC+QLT)  
can consistently map out a non-Fermi liquid regime in the finite-temperature phase diagram of two representative 
spin-fermion models 
involving antiferromagnetic spin density wave (SDW) order and Ising nematic order, respectively. Both models are found to exhibit a fan-like NFL regime above their respective quantum critical points (QCPs), 
which our QMC+QLT approach identifies without any prior knowledge about NFL physics {\sl per se}, nor its rough location in parameter space.
This is accomplished 
by training the respective neural networks to distinguish the quantum states at only {\sl three} parameter points -- one in the  
ordered phase \footnote{For the SDW model, this is technically the quasi-long ranged ordered phase}, one in the low-temperature Fermi liquid on the disordered side of the quantum critical point, and one in the high-temperature regime. Our main results for the observation of a broad NFL regime tracing out a fan above the QCP are illustrated in Fig.~\ref{fig:pds} for the two microscopic 
models exhibiting SDW and nematic order, respectively. In the rest of the paper we first introduce the QMC+QLT approach that allows for this striking observation, its physics-inspired (pre)processing of raw QMC data and its relatively simple neural network architecture. We then discuss our findings from its application to the two microscopic models exhibiting an SDW and nematic QCP, respectively.

 \begin{figure*}
    \centering
	\includegraphics[width=\textwidth]{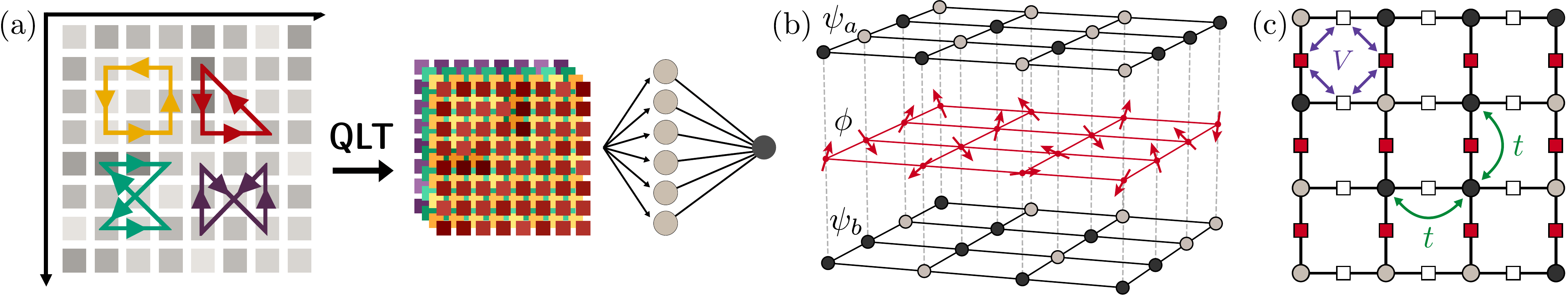}
    \caption{{\bf Architecture of the quantum loop topography approach (a).} A dimensional reduction of the full Green's function data is obtained by only considering correlations along (short) spatial loops. For illustration purposes only four exemplary loops (yellow, red, green, purple) are shown. The resulting quantum loop vector field (colored lattices) are fed into a maximally connected feed-forward neural network. {\bf Illustration of the bilayer lattice models} of Eq.~\eqref{Eq:SDW} featuring a SDW QCP (b) and Eq.~\eqref{Eq:Nematic} hosting an Ising-nematic QCP (c). In the former case (b), two flavors of fermions $\psi_x, \psi_y$ interact with an antiferromagnetic two-component order parameter $\phi$ described by a $\phi^4$-theory. In the latter (c), fermions interact with antiferromagnetically coupled Ising pseudospins (squares) that are situated on the lattice bonds and subject to a transverse field. Based on Figs. 2 and 3 of Refs.~\cite{Bauer2020} and \cite{PhysRevX.6.031028}, respectively.}
    \label{fig:qlt_and_lattice_models}
\end{figure*}

The synergy between machine learning and quantum statistical physics has recently been demonstrated in seminal numerical works performing phase classification tasks \cite{Carrasquilla2017} and quantum state tomography \cite{Carleo2016,Torlai2018}. Of particular relevance for our study here, neural network architectures have recently proven to be capable of identifying established quantum \cite{Broecker2017} and finite-temperature phase transitions \cite{Chng2017} in systems of itinerant electrons. In this work, we significantly expand these approaches to map out 
a non-Fermi liquid regime by employing a preprocessing step using the QLT approach \cite{Zhang2017,Zhang2019}, schematically illustrated in Fig.~\ref{fig:qlt_and_lattice_models}a. In a first step raw samples of the equal-time Green's function, produced in large-scale QMC simulations for a small set of training points chosen to represent the various states in the phase diagram, are preprocessed by extracting triangular and quadrilateral loop correlations. In practice, we limit the length of these ``chained" Green's function products and utilize only a fraction of the available data, namely concentrating on nearest neighbor and next-nearest neighbor information. The resulting quantum loop vector field is then standardized - by subtracting the mean and normalizing to the standard deviation of each feature component - and fed into a fully-connected shallow feed forward artificial neural network (ANN) with a single hidden layer of 
$10$ sigmoid neurons. Afterwards, a supervised training based on stochastic gradient descent and a binary cross-entropy cost function is conducted before the weights and biases of the neural network are frozen. The trained network architecture is then used to classify the quantum states at arbitrary points across the phase diagram, relying only on the fermionic equal-time Green's functions: a choice that renders our approach universally applicable to any intinerant fermion data.

Quantum critical points in itinerant electron systems fall into one of two classes depending on how the Fermi surface changes upon entering the ordered phase. In one class, a gap opens up at select points on the Fermi surface due to an ordering with a finite wave-vector $\mathbf{Q}$. Density waves such as spin-density wave and charge-density waves belong to this class. In the other class, no gap opens anywhere but most of the Fermi surface, if not all, is affected due to a uniform, $\mathbf{Q}=0$ ordering. Nematic order and ferromagnetic order belong to this latter class. To get a comprehensive view, we study representative examples from both classes. 
As a prototype for a system with finite-$\mathbf{Q}$ quantum critical fluctuations, 
we consider
antiferromagnetic spin-density wave fluctuations
in a sign-problem free \cite{Berg2012, Wu2005} spin-fermion system, see Fig.~\ref{fig:qlt_and_lattice_models}(b), that has been investigated by some of us in extensive, numerically exact determinant quantum Monte Carlo (DQMC) studies~\cite{Schattner2016, Gerlach2017}. The action of the two-dimensional lattice model is given by $S = S_\psi + S_\varphi + S_\lambda$, with
\begin{eqnarray}
	\label{Eq:SDW}
	S_\psi &=&  \int_{\tau, \mathbf{k}} \sum_{s, \alpha} \psi_{\alpha \mathbf{k}s}^\dagger \left(\partial_\tau + \epsilon_{\alpha\mathbf{k}s} - \mu\right) \psi_{\alpha \mathbf{k}s} \nonumber\\
	S_\phi &=& \int_{\tau,\mathbf{r}} \left[ \frac{r}{2}\phi^2 + \frac{1}{2c^2} \left(\partial_\tau \phi\right)^2 + \left(\nabla \phi \right)^2 + \frac{u}{4} \phi^4\right]\\
	S_\lambda &=& \lambda \int_{\tau, \mathbf{r}} e^{i \mathbf{Q}\cdot \mathbf{r_i}} \phi_\mathbf{r} \cdot \psi_{a\mathbf{r}s}^\dagger \vec{\sigma}_{ss'}  \psi_{b\mathbf{r}s'} + \textrm{h.c.}, \nonumber
\end{eqnarray}
where $\alpha=a,b$ is a fermion flavor index and $s=\uparrow, \downarrow$ denotes spin. $S_\psi$ describes the free kinetics of two flavors of spin-1/2 fermions $\psi_{\alpha \mathbf{r} s}$ with energy dispersion $\epsilon_{\alpha \mathbf{k} s}$ situated on a square lattice. The antiferromagnetic order parameter $\phi$ is of easy-plane character and is governed by an $O(2)$ symmetric $\phi^4$-theory. The contribution $S_\lambda$ is a Yukawa-like spin-density coupling with an ordering wave vector $\mathbf{Q} = (\pi,\pi)$, which connects different scattering hot spots on the Fermi surface. As specific model parameters, we choose $\lambda=1.5$, $c=3$ and $u=1$, which puts the QCP at a critical coupling $r_c = 0.62$ \cite{Gerlach2017}, masked by the formation of a superconducting dome, as indicated in the lower panel of Fig.~\ref{fig:pds}(a). 

Turning to the numerical analysis of this model and to contrast our QLT approach with a traditional QMC investigation, it is instructive to briefly discuss the results obtained for this model in such a conventional approach~\cite{Gerlach2017}. To locate the SDW phase transition of model~\eqref{Eq:SDW} a careful finite-size scaling analysis has previously been employed for the antiferromagnetic spin-spin correlations: Given that the transition (for an $O(2)$ symmetric order parameter) is of Berezinskii–Kosterlitz–Thouless type, the phase boundary of the ordered state can be identified \cite{Gerlach2017} by tracking its universal critical exponent $\eta = 1/4$. Tracing the transition line down to $T=0$ they obtain an estimate for the QCP position of about $r_c = 0.62$ \cite{Gerlach2017}. Superconductivity,
shown as a dashed line in Fig.~\ref{fig:pds}(a), 
is established by measuring the superfluid density \cite{Scalapino1993}.  
The standard approach to investigating the fate of the fermions in the vicinity of this QCP (above the superconducting dome) is the extraction of the Matsubara self-energy from the imaginary time-displaced Green's function. For model~\eqref{Eq:SDW}, the self-energy at the hot spots is found to be finite and only weakly dependent on Matsubara frequency. This is contrary to the Fermi liquid prediction of a quadratic frequency dependence of the self-energy, and indicates a loss of coherence~\cite{Gerlach2017}.
Consequently, the quasiparticle weight at the hot spots drops significantly near the QCP \cite{Gerlach2017}. While this method for detecting a novel non-Fermi liquid state is theoretically appealing and numerically exact, it is associated with considerable computational cost: the calculation of time-displaced Green's functions. Importantly, this cost is incurred for every parameter point of the phase diagram. For this reason, in spite of the DQMC simulations of Ref.~\cite{Gerlach2017} having a scope of $O(10)$ million CPU hours on modern supercomputers, non-Fermi liquid behavior could only be established at a few discrete points, for example $r=0.7, T=0.05$. Mapping out an extended quantum critical region has so far been out of reach.

As we show in Fig~\ref{fig:pds}(a) our shallow fully connected neural network learns the full two-dimensional phase diagram of the model of Eq.~\eqref{Eq:SDW} when trained with  3200 input vectors from each of the three representative points. With just three points to anchor the phase diagram, it is particularly remarkable that the non-Fermi liquid state (supported by the high temperature training point) extends to the lowest temperatures.  
Moreover, the formation of a quantum critical fan recognized by the neural network shows the steep rise of the NFL-FL crossover temperature away from the quantum critical point without explicit prior knowledge. 
The narrowing of the quantum critical fan zooming into the actual quantum critical point is a highly non-trivial feature learned by the network.
Even more remarkable is the robustness of the phase diagram against the choice of specific training point locations (see SM section IIA Figure 2). This remarkable robustness should be contrasted with neural networks trained on snapshots of classical order parameters, which require training points also in the immediate vicinity of the phase boundary \cite{Carrasquilla2017}, or otherwise fail.

\begin{figure}
    \centering
\includegraphics[width=.5\textwidth, trim={0cm 0cm 0 0cm}, clip]{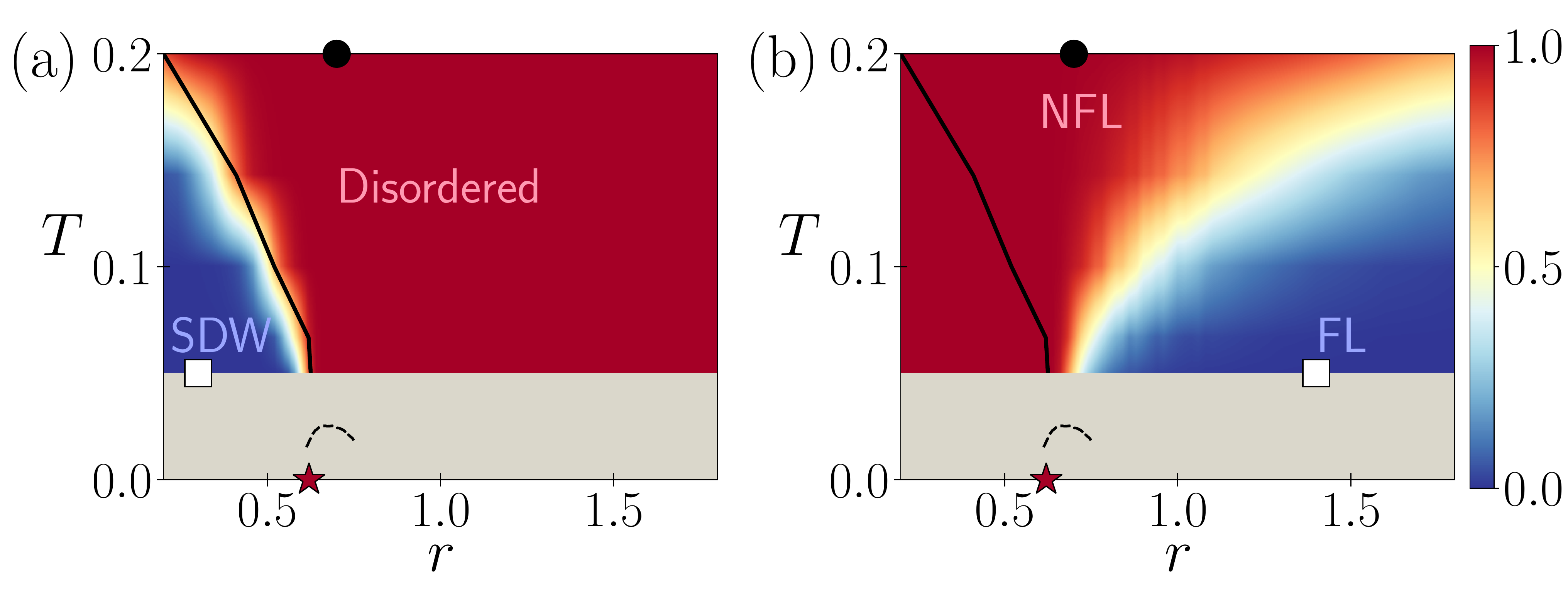}
    \caption{{\bf Binary classification for the SDW model.} (a) The neural network is trained to distinguish the quasi-long range ordered phase (training point at the white box, $r=0.3,T=0.05$) from the non-Fermi liquid regime (the black circle, $r=0.7,T=0.2$). (b) The neural network is trained to distinguish the disordered Fermi liquid regime (training point at the white box, $r=1.1,T=0.05$) from the non-Fermi liquid regime (the black circle, as in (a)).}
    \label{fig:2ptsdw}
\end{figure}

To further investigate the neural network's learning of the NFL region and the quantum critical fan, we explore a setup where the  ANN learns only about the existence of the SDW phase and the FL state by training with only {\sl two sets} of training data for each case. 
Even though QLT was originally developed as a probe of transport \cite{Zhang2019}, a binary classification using the SDW 
and disordered phases robustly captures  almost the entire phase boundary of the SDW state, as shown in  Fig~\ref{fig:2ptsdw}(a).
This level of performance without explicit reference to the order parameter comes somewhat surprising. We speculate that the gap opened by the SDW order 
in the vicinity of only four points on the Fermi surface may nonetheless be
robustly detectable in the (local) 
QLT input. 
A binary classification targeting the FL  and disordered states, shown in  Fig~\ref{fig:2ptsdw}(b), indicates that the NFL-FL cross-over can also be independently learned, even without referencing the ordered phase.

As a prototype of a quantum critical point to a uniform ($\mathbf{Q}=0$) order, we consider a sign-problem free lattice model for Ising nematic quantum criticality~\cite{PhysRevX.6.031028,Lederer4905}. As shown in Fig.~\ref{fig:qlt_and_lattice_models}(c),
the model's degrees of freedom are 
fermions $c_{i,\sigma}$ that live on the sites $i$ of a square lattice, 
and pseudospins that live on the nearest neighbor bonds $\left<i,j\right>$ coupling to the bond charge density of fermions. The Hamiltonian is $H = H_{f} + H_{b} + H_{int}$, where
\begin{align}
\label{Eq:Nematic}
    H_{f} &= -t\sum_{\left<i,j\right>,\sigma}c_{i\sigma}^{\dagger}c_{j\sigma}-\mu \sum_{i\sigma}c_{i\sigma}^{\dagger}c_{i\sigma} \nonumber \\
    H_{b} &= V \sum_{\left<\left<i,j\right>;\left<k,l\right>\right>}\tau^{z}_{i,j}\tau^{z}_{k,l}-h\sum_{\left<i,j\right>}\tau^{x}_{i,j}\\
    H_{int} &= \alpha t\sum_{\left<i,j\right>,\sigma}\tau^{z}_{i,j}c_{i\sigma}^{\dagger}c_{j\sigma}. \nonumber
\end{align}
Here,  fermion hopping is  considered 
only between nearest neighbor sites, and the `antiferromagnetic' interaction $V$ between pseudospins on nearest neighbor bonds drives the nematic order: when the $z$ components of pseudospins on horizontal bonds, i.e. the white squares in Fig.~\ref{fig:qlt_and_lattice_models}(c), differ from those on their neighboring vertical bonds (red squares), the effective hopping of fermions becomes anisotropic because of the spin-fermion coupling (third line of Eq.~\eqref{Eq:Nematic}).     
The ``transverse field'' $h$ frustrates the ordering tendency and introduces dynamics.

The traditional approach of investigating the phase diagram   
of model \eqref{Eq:Nematic} starts by determining the phase boundary of the nematic order. Because the nematic order parameter is discrete, any given simulation will be kinetically trapped in one of the two degenerate minima, making QMC  snapshots incompatible amongst different parallel branches. 
Traditionally this issue is avoided by turning to the order parameter correlation function (i.e. the nematic susceptibility) and its finite-size scaling \cite{Binder} for the determination of the ordered phase boundary, as it was done in Ref.~\cite{Lederer4905}  using the known 2D Ising critical exponents. 
While notionally rigorous, this technique requires simulations of numerous system sizes, and involves some guesswork in determining the phase transition using data collapse. 

Here, we take an alternate approach of 
 a ``cold start" \footnote{ The ``cold start'' initializes the nematic pseudospins in one of the classical ground states of $H_b$ to bias the Monte Carlo kinetics and ensure that the simulations converge to the vicinity of a single local minimum.} for $h$ below that of the non-Fermi liquid training point ($h=2.7$). This way we bypass the computational cost of finite size scaling and evaluation of higher order fermion 
 correlation functions. Instead we work only with the equal-time Green's functions from simulations on a single system size.
Since nearest neighbor Green's functions are conjugate to 
the order parameter, we provide nearest neighbor Green's function data to the feature vector in addition to QLT. 
 As shown in Fig.~\ref{fig:pds}, the ANN learns the nematic phase boundary  
in remarkable agreement with the conventional analysis~\cite{Lederer4905}, indicated by the solid line in Fig.~\ref{fig:pds}, down to the lowest temperatures.

Outside of the nematic phase, a broad superconducting dome forms \cite{Lederer4905} with a relatively high transition temperature, indicated by the dashed line in Fig.~\ref{fig:pds}(b). This is accompanied by a violation of Fermi liquid predictions in the quantum critical region, indicated by a self-energy whose imaginary part becomes large and, as in the SDW model, frequency independent.
 There is also evidence of non-Fermi liquid transport from various proxies for the
 DC resistivity, but these are all subject to the considerable ambiguities of all known forms of analytic continuation.

The efficient learning of the full two dimensional phase diagram using only three training points deep in the states of interest in Fig.~\ref{fig:pds} is particularly remarkable 
given that all three states are gapless. (The Ising nematic order leaves the Fermi surface gapless with well-defined quasi-particles, only introducing anisotropy to the Fermi surface.)  
It is thus the subtler changes in the Green's function among the three gapless phases, that the ANN learns in determining the the full phase diagram and indeed recognizes the NFL region down to the lowest temperatures (and despite the NFL training point being located at the highest temperature). Moreover, the phase diagram is once again robust against the choice of the NFL training point (see SM section III).

\begin{figure} [t]
    \centering
    \includegraphics[width=\linewidth, trim={0 0cm 0 0cm}, clip]{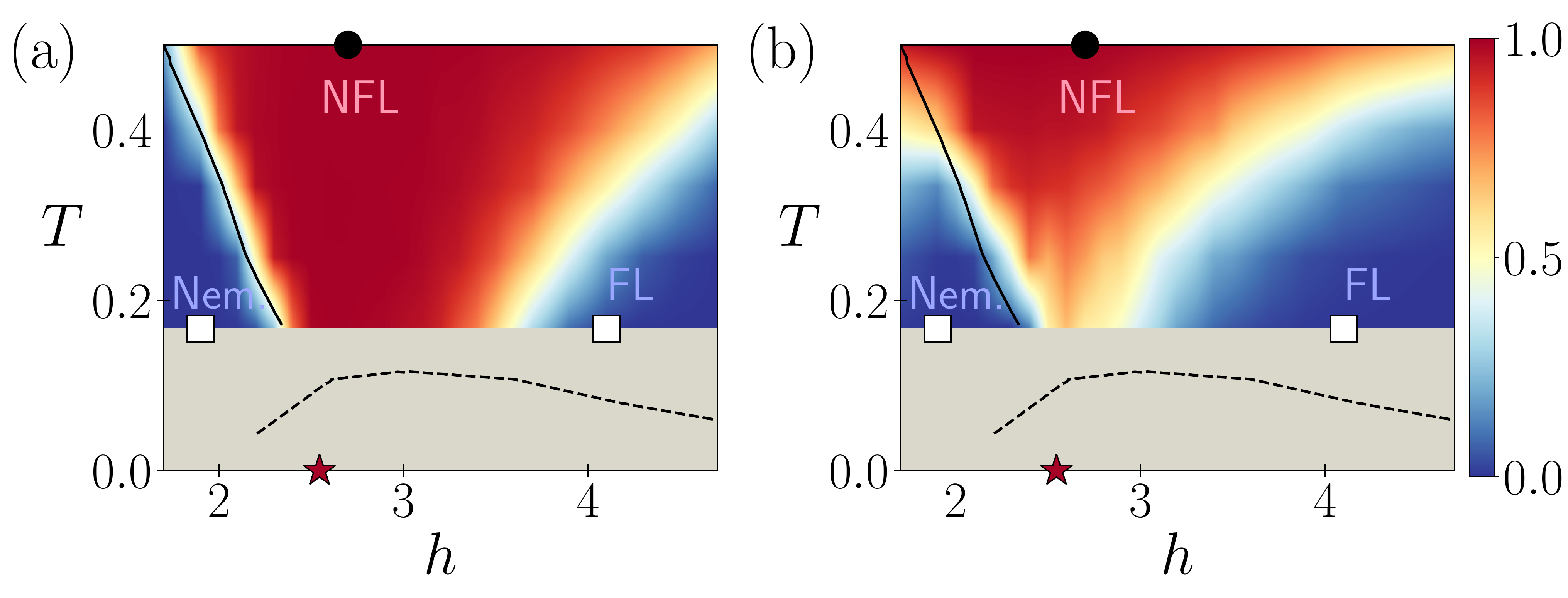}
    \caption{\textbf{Comparison of the feature vectors for the nematic model.} Analogous to Figure 1, the predictions of the non-Fermi liquid region are shown for a neural network trained on (a) only QLT data and (b) only the nearest neighbor Green's functions data for the nematic model. The training points are identical to those used previously, with white boxes for the Fermi liquid ( $h=1.9,T=0.17$ and $h=4.1,T=0.17$) and black circles for the non-Fermi liquid ($h=2.7,T=0.5$).
        \label{fig:nem_preprocess}
    }
\end{figure}

To further investigate the robustness of the ANN learning of NFL phenomena we also looked into how the ANN learning depends on the feature vector input. 
For Fig.~\ref{fig:pds}, we used QLT together with the nearest neighbor Green's function as input to the ANN. 
Analogous phase diagrams 
using QLT input alone and the nearest-neighbor Green's functions alone are shown in Fig.~\ref{fig:nem_preprocess} (a) and (b), respectively. The comparison shows that the overall features of the NFL regions are consistent across different choices of the input features. 
QLT data allow the ANN to recognize a definitive NFL region extending down to the lowest temperatures from the highest temperature training point, while the nearest neighbor Green's function renders a NFL region that narrows with weakening confidence upon cooling. On the other hand, the nearest neighbor Green's functions offer more robustness in the phase classification against shifts in Fermi liquid training points (see Supplemental Materials section III for further details).

To summarize, we took a data-science approach to the vast volume of war data generated by QMC simulations of quantum critical phenomena of itinerant fermions coupled to antiferromagnetic spin-density wave or Ising-nematic order.  By simply providing the equal-time single-particle Green's function data, processed using a QLT machine learning approach that is designed to target transport, we obtained detailed features of the full phase diagrams, including the formation of NFL physics in a quantum critical fan above the QCP, from the raw data for both models. Our analysis relied  on the simulations for a single system size only and just three training points deep in the respective phases, but proofs remarkably consistent with the traditionally obtained phase boundaries for the ordered phases.  
But most notably, the NFL region is clearly and robustly identified directly from the equal-time data.

Our results prove that it is indeed possible to efficiently extract the information relevant for identifying NFL physics encoded in the 
equal-time, position-space Green's function data directly. 
Indeed, the full exploration of the quantum critical region from an exact simulation of SDW model (which in a conventional analysis turned out to be elusively expensive) was made possible for the first time using this QMC+QLT approach described in this manuscript. The obtained phase diagram clearly reveals the quantum critical point at $T=0$, unknown to the ANN, to be the singlular anchor of the NFL regime -- perhaps the most subtle and mysterious state that itinerant fermions can form. The simplicity and the robustness of our approach combined with its effectiveness in detecting this subtle state imply that data scientific approaches can enable discoveries from the data readily accessible to QMC simulations in future explorations.

{\it Acknowledgements.--} 
SL and E-AK acknowledge the support from the U.S. Department of Energy, Office of Basic Energy Sciences, Division of Materials Science and Engineering under Award DE-SC0018946.
The Cologne group acknowledges partial support from the Deutsche Forschungsgemeinschaft (DFG, German Research Foundation) -- Projektnummer 277101999 -- TRR 183 (project B01).
The numerical simulations were performed on the JUWELS cluster at FZ J\"ulich and the CHEOPS cluster at RRZK Cologne.

\bibliography{refs}

\begin{thebibliography}{42}%
\makeatletter
\providecommand \@ifxundefined [1]{%
 \@ifx{#1\undefined}
}%
\providecommand \@ifnum [1]{%
 \ifnum #1\expandafter \@firstoftwo
 \else \expandafter \@secondoftwo
 \fi
}%
\providecommand \@ifx [1]{%
 \ifx #1\expandafter \@firstoftwo
 \else \expandafter \@secondoftwo
 \fi
}%
\providecommand \natexlab [1]{#1}%
\providecommand \enquote  [1]{``#1''}%
\providecommand \bibnamefont  [1]{#1}%
\providecommand \bibfnamefont [1]{#1}%
\providecommand \citenamefont [1]{#1}%
\providecommand \href@noop [0]{\@secondoftwo}%
\providecommand \href [0]{\begingroup \@sanitize@url \@href}%
\providecommand \@href[1]{\@@startlink{#1}\@@href}%
\providecommand \@@href[1]{\endgroup#1\@@endlink}%
\providecommand \@sanitize@url [0]{\catcode `\\12\catcode `\$12\catcode
  `\&12\catcode `\#12\catcode `\^12\catcode `\_12\catcode `\%12\relax}%
\providecommand \@@startlink[1]{}%
\providecommand \@@endlink[0]{}%
\providecommand \url  [0]{\begingroup\@sanitize@url \@url }%
\providecommand \@url [1]{\endgroup\@href {#1}{\urlprefix }}%
\providecommand \urlprefix  [0]{URL }%
\providecommand \Eprint [0]{\href }%
\providecommand \doibase [0]{http://dx.doi.org/}%
\providecommand \selectlanguage [0]{\@gobble}%
\providecommand \bibinfo  [0]{\@secondoftwo}%
\providecommand \bibfield  [0]{\@secondoftwo}%
\providecommand \translation [1]{[#1]}%
\providecommand \BibitemOpen [0]{}%
\providecommand \bibitemStop [0]{}%
\providecommand \bibitemNoStop [0]{.\EOS\space}%
\providecommand \EOS [0]{\spacefactor3000\relax}%
\providecommand \BibitemShut  [1]{\csname bibitem#1\endcsname}%
\let\auto@bib@innerbib\@empty
\bibitem [{\citenamefont {Haldane}(1994)}]{haldane1994}%
  \BibitemOpen
  \bibfield  {author} {\bibinfo {author} {\bibfnamefont {F.~D.~M.}\
  \bibnamefont {Haldane}},\ }in\ \href@noop {} {\emph {\bibinfo {booktitle}
  {{Proceedings of the International School of Physics "Enrico Fermi", Course
  CXXI "Perspectives in Many-Particle Physics"}}}},\ \bibinfo {editor} {edited
  by\ \bibinfo {editor} {\bibfnamefont {R.A.}\ \bibnamefont {Broglia}}\ and\
  \bibinfo {editor} {\bibfnamefont {J.R.}\ \bibnamefont {Schrieffer}}}\
  (\bibinfo {year} {1994})\ pp.\ \bibinfo {pages} {5--29},\ \Eprint
  {http://arxiv.org/abs/cond-mat/0505529} {arXiv:cond-mat/0505529} \BibitemShut
  {NoStop}%
\bibitem [{\citenamefont {Ong}\ and\ \citenamefont
  {Bhatt}(2001)}]{ong2001more}%
  \BibitemOpen
  \bibfield  {author} {\bibinfo {author} {\bibfnamefont {Nai-Phuan}\
  \bibnamefont {Ong}}\ and\ \bibinfo {author} {\bibfnamefont {Ravin}\
  \bibnamefont {Bhatt}},\ }\bibfield  {title} {\enquote {\bibinfo {title}
  {{More is Different}},}\ }\href@noop {} {\bibfield  {journal} {\bibinfo
  {journal} {More is Different by Nai-Phuan Ong and Ravin Bhatt. Princeton
  University Press}\ } (\bibinfo {year} {2001})}\BibitemShut {NoStop}%
\bibitem [{\citenamefont {L\"ohneysen}\ \emph {et~al.}(2007)\citenamefont
  {L\"ohneysen}, \citenamefont {Rosch}, \citenamefont {Vojta},\ and\
  \citenamefont {W\"olfle}}]{lohneysen2007}%
  \BibitemOpen
  \bibfield  {author} {\bibinfo {author} {\bibfnamefont {Hilbert~v.}\
  \bibnamefont {L\"ohneysen}}, \bibinfo {author} {\bibfnamefont {Achim}\
  \bibnamefont {Rosch}}, \bibinfo {author} {\bibfnamefont {Matthias}\
  \bibnamefont {Vojta}}, \ and\ \bibinfo {author} {\bibfnamefont {Peter}\
  \bibnamefont {W\"olfle}},\ }\bibfield  {title} {\enquote {\bibinfo {title}
  {Fermi-liquid instabilities at magnetic quantum phase transitions},}\ }\href
  {\doibase 10.1103/RevModPhys.79.1015} {\bibfield  {journal} {\bibinfo
  {journal} {Rev. Mod. Phys.}\ }\textbf {\bibinfo {volume} {79}},\ \bibinfo
  {pages} {1015--1075} (\bibinfo {year} {2007})}\BibitemShut {NoStop}%
\bibitem [{\citenamefont {Sachdev}\ and\ \citenamefont
  {Keimer}(2011)}]{SachdevKeimer2011}%
  \BibitemOpen
  \bibfield  {author} {\bibinfo {author} {\bibfnamefont {Subir}\ \bibnamefont
  {Sachdev}}\ and\ \bibinfo {author} {\bibfnamefont {Bernhard}\ \bibnamefont
  {Keimer}},\ }\bibfield  {title} {\enquote {\bibinfo {title} {{Quantum
  criticality}},}\ }\href {\doibase 10.1063/1.3554314} {\bibfield  {journal}
  {\bibinfo  {journal} {Physics Today}\ }\textbf {\bibinfo {volume} {64}},\
  \bibinfo {pages} {29--35} (\bibinfo {year} {2011})}\BibitemShut {NoStop}%
\bibitem [{\citenamefont {Rosch}(1999)}]{PhysRevLett.82.4280}%
  \BibitemOpen
  \bibfield  {author} {\bibinfo {author} {\bibfnamefont {A.}~\bibnamefont
  {Rosch}},\ }\bibfield  {title} {\enquote {\bibinfo {title} {Interplay of
  disorder and spin fluctuations in the resistivity near a quantum critical
  point},}\ }\href {\doibase 10.1103/PhysRevLett.82.4280} {\bibfield  {journal}
  {\bibinfo  {journal} {Phys. Rev. Lett.}\ }\textbf {\bibinfo {volume} {82}},\
  \bibinfo {pages} {4280--4283} (\bibinfo {year} {1999})}\BibitemShut {NoStop}%
\bibitem [{\citenamefont {Dell'Anna}\ and\ \citenamefont
  {Metzner}(2007)}]{PhysRevLett.98.136402}%
  \BibitemOpen
  \bibfield  {author} {\bibinfo {author} {\bibfnamefont {Luca}\ \bibnamefont
  {Dell'Anna}}\ and\ \bibinfo {author} {\bibfnamefont {Walter}\ \bibnamefont
  {Metzner}},\ }\bibfield  {title} {\enquote {\bibinfo {title} {Electrical
  resistivity near pomeranchuk instability in two dimensions},}\ }\href
  {\doibase 10.1103/PhysRevLett.98.136402} {\bibfield  {journal} {\bibinfo
  {journal} {Phys. Rev. Lett.}\ }\textbf {\bibinfo {volume} {98}},\ \bibinfo
  {pages} {136402} (\bibinfo {year} {2007})}\BibitemShut {NoStop}%
\bibitem [{\citenamefont {Maslov}\ \emph {et~al.}(2011)\citenamefont {Maslov},
  \citenamefont {Yudson},\ and\ \citenamefont
  {Chubukov}}]{PhysRevLett.106.106403}%
  \BibitemOpen
  \bibfield  {author} {\bibinfo {author} {\bibfnamefont {Dmitrii~L.}\
  \bibnamefont {Maslov}}, \bibinfo {author} {\bibfnamefont {Vladimir~I.}\
  \bibnamefont {Yudson}}, \ and\ \bibinfo {author} {\bibfnamefont {Andrey~V.}\
  \bibnamefont {Chubukov}},\ }\bibfield  {title} {\enquote {\bibinfo {title}
  {Resistivity of a non-galilean--invariant fermi liquid near pomeranchuk
  quantum criticality},}\ }\href {\doibase 10.1103/PhysRevLett.106.106403}
  {\bibfield  {journal} {\bibinfo  {journal} {Phys. Rev. Lett.}\ }\textbf
  {\bibinfo {volume} {106}},\ \bibinfo {pages} {106403} (\bibinfo {year}
  {2011})}\BibitemShut {NoStop}%
\bibitem [{\citenamefont {Hartnoll}\ \emph {et~al.}(2014)\citenamefont
  {Hartnoll}, \citenamefont {Mahajan}, \citenamefont {Punk},\ and\
  \citenamefont {Sachdev}}]{PhysRevB.89.155130}%
  \BibitemOpen
  \bibfield  {author} {\bibinfo {author} {\bibfnamefont {Sean~A.}\ \bibnamefont
  {Hartnoll}}, \bibinfo {author} {\bibfnamefont {Raghu}\ \bibnamefont
  {Mahajan}}, \bibinfo {author} {\bibfnamefont {Matthias}\ \bibnamefont
  {Punk}}, \ and\ \bibinfo {author} {\bibfnamefont {Subir}\ \bibnamefont
  {Sachdev}},\ }\bibfield  {title} {\enquote {\bibinfo {title} {Transport near
  the ising-nematic quantum critical point of metals in two dimensions},}\
  }\href {\doibase 10.1103/PhysRevB.89.155130} {\bibfield  {journal} {\bibinfo
  {journal} {Phys. Rev. B}\ }\textbf {\bibinfo {volume} {89}},\ \bibinfo
  {pages} {155130} (\bibinfo {year} {2014})}\BibitemShut {NoStop}%
\bibitem [{\citenamefont {Patel}\ and\ \citenamefont
  {Sachdev}(2014)}]{PhysRevB.90.165146}%
  \BibitemOpen
  \bibfield  {author} {\bibinfo {author} {\bibfnamefont {Aavishkar~A.}\
  \bibnamefont {Patel}}\ and\ \bibinfo {author} {\bibfnamefont {Subir}\
  \bibnamefont {Sachdev}},\ }\bibfield  {title} {\enquote {\bibinfo {title} {dc
  resistivity at the onset of spin density wave order in two-dimensional
  metals},}\ }\href {\doibase 10.1103/PhysRevB.90.165146} {\bibfield  {journal}
  {\bibinfo  {journal} {Phys. Rev. B}\ }\textbf {\bibinfo {volume} {90}},\
  \bibinfo {pages} {165146} (\bibinfo {year} {2014})}\BibitemShut {NoStop}%
\bibitem [{\citenamefont {Wang}\ and\ \citenamefont
  {Berg}(2019)}]{PhysRevB.99.235136}%
  \BibitemOpen
  \bibfield  {author} {\bibinfo {author} {\bibfnamefont {Xiaoyu}\ \bibnamefont
  {Wang}}\ and\ \bibinfo {author} {\bibfnamefont {Erez}\ \bibnamefont {Berg}},\
  }\bibfield  {title} {\enquote {\bibinfo {title} {Scattering mechanisms and
  electrical transport near an ising nematic quantum critical point},}\ }\href
  {\doibase 10.1103/PhysRevB.99.235136} {\bibfield  {journal} {\bibinfo
  {journal} {Phys. Rev. B}\ }\textbf {\bibinfo {volume} {99}},\ \bibinfo
  {pages} {235136} (\bibinfo {year} {2019})}\BibitemShut {NoStop}%
\bibitem [{\citenamefont {de~Carvalho}\ and\ \citenamefont
  {Fernandes}(2019)}]{PhysRevB.100.115103}%
  \BibitemOpen
  \bibfield  {author} {\bibinfo {author} {\bibfnamefont {V.~S.}\ \bibnamefont
  {de~Carvalho}}\ and\ \bibinfo {author} {\bibfnamefont {R.~M.}\ \bibnamefont
  {Fernandes}},\ }\bibfield  {title} {\enquote {\bibinfo {title} {Resistivity
  near a nematic quantum critical point: Impact of acoustic phonons},}\ }\href
  {\doibase 10.1103/PhysRevB.100.115103} {\bibfield  {journal} {\bibinfo
  {journal} {Phys. Rev. B}\ }\textbf {\bibinfo {volume} {100}},\ \bibinfo
  {pages} {115103} (\bibinfo {year} {2019})}\BibitemShut {NoStop}%
\bibitem [{\citenamefont {Vieira}\ \emph {et~al.}(2020)\citenamefont {Vieira},
  \citenamefont {[de Carvalho]},\ and\ \citenamefont
  {Freire}}]{VIEIRA2020168230}%
  \BibitemOpen
  \bibfield  {author} {\bibinfo {author} {\bibfnamefont {Lucas~E.}\
  \bibnamefont {Vieira}}, \bibinfo {author} {\bibfnamefont {Vanuildo~S.}\
  \bibnamefont {[de Carvalho]}}, \ and\ \bibinfo {author} {\bibfnamefont
  {Hermann}\ \bibnamefont {Freire}},\ }\bibfield  {title} {\enquote {\bibinfo
  {title} {Dc resistivity near a nematic quantum critical point: Effects of
  weak disorder and acoustic phonons},}\ }\href {\doibase
  https://doi.org/10.1016/j.aop.2020.168230} {\bibfield  {journal} {\bibinfo
  {journal} {Annals of Physics}\ }\textbf {\bibinfo {volume} {419}},\ \bibinfo
  {pages} {168230} (\bibinfo {year} {2020})}\BibitemShut {NoStop}%
\bibitem [{\citenamefont {Loh}\ \emph {et~al.}(1990)\citenamefont {Loh},
  \citenamefont {Gubernatis}, \citenamefont {Scalettar}, \citenamefont {White},
  \citenamefont {Scalapino},\ and\ \citenamefont {Sugar}}]{Loh1990}%
  \BibitemOpen
  \bibfield  {author} {\bibinfo {author} {\bibfnamefont {E.~Y.}\ \bibnamefont
  {Loh}}, \bibinfo {author} {\bibfnamefont {J.~E.}\ \bibnamefont {Gubernatis}},
  \bibinfo {author} {\bibfnamefont {R.~T.}\ \bibnamefont {Scalettar}}, \bibinfo
  {author} {\bibfnamefont {S.~R.}\ \bibnamefont {White}}, \bibinfo {author}
  {\bibfnamefont {D.~J.}\ \bibnamefont {Scalapino}}, \ and\ \bibinfo {author}
  {\bibfnamefont {R.~L.}\ \bibnamefont {Sugar}},\ }\bibfield  {title} {\enquote
  {\bibinfo {title} {{Sign problem in the numerical simulation of many-electron
  systems}},}\ }\href {\doibase 10.1103/PhysRevB.41.9301} {\bibfield  {journal}
  {\bibinfo  {journal} {Phys. Rev. B}\ }\textbf {\bibinfo {volume} {41}},\
  \bibinfo {pages} {9301--9307} (\bibinfo {year} {1990})}\BibitemShut {NoStop}%
\bibitem [{\citenamefont {Gioev}\ and\ \citenamefont
  {Klich}(2006)}]{Klich2006}%
  \BibitemOpen
  \bibfield  {author} {\bibinfo {author} {\bibfnamefont {Dimitri}\ \bibnamefont
  {Gioev}}\ and\ \bibinfo {author} {\bibfnamefont {Israel}\ \bibnamefont
  {Klich}},\ }\bibfield  {title} {\enquote {\bibinfo {title} {{Entanglement
  Entropy of Fermions in Any Dimension and the Widom Conjecture}},}\ }\href
  {\doibase 10.1103/PhysRevLett.96.100503} {\bibfield  {journal} {\bibinfo
  {journal} {Phys. Rev. Lett.}\ }\textbf {\bibinfo {volume} {96}},\ \bibinfo
  {pages} {100503} (\bibinfo {year} {2006})}\BibitemShut {NoStop}%
\bibitem [{\citenamefont {Wolf}(2006)}]{Wolf2006}%
  \BibitemOpen
  \bibfield  {author} {\bibinfo {author} {\bibfnamefont {Michael~M.}\
  \bibnamefont {Wolf}},\ }\bibfield  {title} {\enquote {\bibinfo {title}
  {{Violation of the Entropic Area Law for Fermions}},}\ }\href {\doibase
  10.1103/PhysRevLett.96.010404} {\bibfield  {journal} {\bibinfo  {journal}
  {Phys. Rev. Lett.}\ }\textbf {\bibinfo {volume} {96}},\ \bibinfo {pages}
  {010404} (\bibinfo {year} {2006})}\BibitemShut {NoStop}%
\bibitem [{\citenamefont {Parcollet}\ and\ \citenamefont
  {Georges}(1999)}]{parcollet1999}%
  \BibitemOpen
  \bibfield  {author} {\bibinfo {author} {\bibfnamefont {Olivier}\ \bibnamefont
  {Parcollet}}\ and\ \bibinfo {author} {\bibfnamefont {Antoine}\ \bibnamefont
  {Georges}},\ }\bibfield  {title} {\enquote {\bibinfo {title}
  {{Non-Fermi-liquid regime of a doped Mott insulator}},}\ }\href {\doibase
  10.1103/PhysRevB.59.5341} {\bibfield  {journal} {\bibinfo  {journal} {Phys.
  Rev. B}\ }\textbf {\bibinfo {volume} {59}},\ \bibinfo {pages} {5341--5360}
  (\bibinfo {year} {1999})}\BibitemShut {NoStop}%
\bibitem [{\citenamefont {Cha}\ \emph {et~al.}(2019)\citenamefont {Cha},
  \citenamefont {Patel}, \citenamefont {Gull},\ and\ \citenamefont
  {Kim}}]{cha2019tlinear}%
  \BibitemOpen
  \bibfield  {author} {\bibinfo {author} {\bibfnamefont {Peter}\ \bibnamefont
  {Cha}}, \bibinfo {author} {\bibfnamefont {Aavishkar~A.}\ \bibnamefont
  {Patel}}, \bibinfo {author} {\bibfnamefont {Emanuel}\ \bibnamefont {Gull}}, \
  and\ \bibinfo {author} {\bibfnamefont {Eun-Ah}\ \bibnamefont {Kim}},\
  }\href@noop {} {\enquote {\bibinfo {title} {$t$-linear resistivity in models
  with local self-energy},}\ } (\bibinfo {year} {2019}),\ \Eprint
  {http://arxiv.org/abs/1910.07530} {arXiv:1910.07530 [cond-mat.str-el]}
  \BibitemShut {NoStop}%
\bibitem [{\citenamefont {Cha}\ \emph {et~al.}(2020)\citenamefont {Cha},
  \citenamefont {Wentzell}, \citenamefont {Parcollet}, \citenamefont
  {Georges},\ and\ \citenamefont {Kim}}]{cha2020linear}%
  \BibitemOpen
  \bibfield  {author} {\bibinfo {author} {\bibfnamefont {Peter}\ \bibnamefont
  {Cha}}, \bibinfo {author} {\bibfnamefont {Nils}\ \bibnamefont {Wentzell}},
  \bibinfo {author} {\bibfnamefont {Olivier}\ \bibnamefont {Parcollet}},
  \bibinfo {author} {\bibfnamefont {Antoine}\ \bibnamefont {Georges}}, \ and\
  \bibinfo {author} {\bibfnamefont {Eun-Ah}\ \bibnamefont {Kim}},\ }\bibfield
  {title} {\enquote {\bibinfo {title} {Linear resistivity and sachdev-ye-kitaev
  (syk) spin liquid behavior in a quantum critical metal with spin-$1/2$
  fermions},}\ }\href@noop {} {\bibfield  {journal} {\bibinfo  {journal}
  {arXiv:2002.07181}\ } (\bibinfo {year} {2020})}\BibitemShut {NoStop}%
\bibitem [{\citenamefont {Berg}\ \emph {et~al.}(2012)\citenamefont {Berg},
  \citenamefont {Metlitski},\ and\ \citenamefont {Sachdev}}]{Berg2012}%
  \BibitemOpen
  \bibfield  {author} {\bibinfo {author} {\bibfnamefont {Erez}\ \bibnamefont
  {Berg}}, \bibinfo {author} {\bibfnamefont {Max~A}\ \bibnamefont {Metlitski}},
  \ and\ \bibinfo {author} {\bibfnamefont {Subir}\ \bibnamefont {Sachdev}},\
  }\bibfield  {title} {\enquote {\bibinfo {title} {{Sign-Problem-Free Quantum
  Monte Carlo of the Onset of Antiferromagnetism in Metals}},}\ }\href
  {\doibase 10.1126/science.1227769} {\bibfield  {journal} {\bibinfo  {journal}
  {Science}\ }\textbf {\bibinfo {volume} {338}},\ \bibinfo {pages} {1606--1609}
  (\bibinfo {year} {2012})}\BibitemShut {NoStop}%
\bibitem [{\citenamefont {Berg}\ \emph {et~al.}(2019)\citenamefont {Berg},
  \citenamefont {Lederer}, \citenamefont {Schattner},\ and\ \citenamefont
  {Trebst}}]{Berg2019}%
  \BibitemOpen
  \bibfield  {author} {\bibinfo {author} {\bibfnamefont {Erez}\ \bibnamefont
  {Berg}}, \bibinfo {author} {\bibfnamefont {Samuel}\ \bibnamefont {Lederer}},
  \bibinfo {author} {\bibfnamefont {Yoni}\ \bibnamefont {Schattner}}, \ and\
  \bibinfo {author} {\bibfnamefont {Simon}\ \bibnamefont {Trebst}},\ }\bibfield
   {title} {\enquote {\bibinfo {title} {{Monte Carlo Studies of Quantum
  Critical Metals}},}\ }\href {\doibase
  10.1146/annurev-conmatphys-031218-013339} {\bibfield  {journal} {\bibinfo
  {journal} {Annual Review of Condensed Matter Physics}\ }\textbf {\bibinfo
  {volume} {10}},\ \bibinfo {pages} {63--84} (\bibinfo {year}
  {2019})}\BibitemShut {NoStop}%
\bibitem [{\citenamefont {Xu}\ \emph {et~al.}(2019)\citenamefont {Xu},
  \citenamefont {Liu}, \citenamefont {Pan}, \citenamefont {Qi}, \citenamefont
  {Sun},\ and\ \citenamefont {Meng}}]{Xu_2019}%
  \BibitemOpen
  \bibfield  {author} {\bibinfo {author} {\bibfnamefont {Xiao~Yan}\
  \bibnamefont {Xu}}, \bibinfo {author} {\bibfnamefont {Zi~Hong}\ \bibnamefont
  {Liu}}, \bibinfo {author} {\bibfnamefont {Gaopei}\ \bibnamefont {Pan}},
  \bibinfo {author} {\bibfnamefont {Yang}\ \bibnamefont {Qi}}, \bibinfo
  {author} {\bibfnamefont {Kai}\ \bibnamefont {Sun}}, \ and\ \bibinfo {author}
  {\bibfnamefont {Zi~Yang}\ \bibnamefont {Meng}},\ }\bibfield  {title}
  {\enquote {\bibinfo {title} {Revealing fermionic quantum criticality from new
  monte carlo techniques},}\ }\href {\doibase 10.1088/1361-648x/ab3295}
  {\bibfield  {journal} {\bibinfo  {journal} {Journal of Physics: Condensed
  Matter}\ }\textbf {\bibinfo {volume} {31}},\ \bibinfo {pages} {463001}
  (\bibinfo {year} {2019})}\BibitemShut {NoStop}%
\bibitem [{\citenamefont {Schattner}\ \emph
  {et~al.}(2016{\natexlab{a}})\citenamefont {Schattner}, \citenamefont
  {Gerlach}, \citenamefont {Trebst},\ and\ \citenamefont
  {Berg}}]{Schattner2016}%
  \BibitemOpen
  \bibfield  {author} {\bibinfo {author} {\bibfnamefont {Yoni}\ \bibnamefont
  {Schattner}}, \bibinfo {author} {\bibfnamefont {Max~H.}\ \bibnamefont
  {Gerlach}}, \bibinfo {author} {\bibfnamefont {Simon}\ \bibnamefont {Trebst}},
  \ and\ \bibinfo {author} {\bibfnamefont {Erez}\ \bibnamefont {Berg}},\
  }\bibfield  {title} {\enquote {\bibinfo {title} {{Competing Orders in a
  Nearly Antiferromagnetic Metal}},}\ }\href {\doibase
  10.1103/PhysRevLett.117.097002} {\bibfield  {journal} {\bibinfo  {journal}
  {Phys. Rev. Lett.}\ }\textbf {\bibinfo {volume} {117}},\ \bibinfo {pages}
  {097002} (\bibinfo {year} {2016}{\natexlab{a}})}\BibitemShut {NoStop}%
\bibitem [{\citenamefont {Gerlach}\ \emph {et~al.}(2017)\citenamefont
  {Gerlach}, \citenamefont {Schattner}, \citenamefont {Berg},\ and\
  \citenamefont {Trebst}}]{Gerlach2017}%
  \BibitemOpen
  \bibfield  {author} {\bibinfo {author} {\bibfnamefont {Max~H.}\ \bibnamefont
  {Gerlach}}, \bibinfo {author} {\bibfnamefont {Yoni}\ \bibnamefont
  {Schattner}}, \bibinfo {author} {\bibfnamefont {Erez}\ \bibnamefont {Berg}},
  \ and\ \bibinfo {author} {\bibfnamefont {Simon}\ \bibnamefont {Trebst}},\
  }\bibfield  {title} {\enquote {\bibinfo {title} {{Quantum critical properties
  of a metallic spin-density-wave transition}},}\ }\href {\doibase
  10.1103/PhysRevB.95.035124} {\bibfield  {journal} {\bibinfo  {journal} {Phys.
  Rev. B}\ }\textbf {\bibinfo {volume} {95}},\ \bibinfo {pages} {035124}
  (\bibinfo {year} {2017})}\BibitemShut {NoStop}%
\bibitem [{\citenamefont {Liu}\ \emph {et~al.}(2019)\citenamefont {Liu},
  \citenamefont {Pan}, \citenamefont {Xu}, \citenamefont {Sun},\ and\
  \citenamefont {Meng}}]{Liu2018}%
  \BibitemOpen
  \bibfield  {author} {\bibinfo {author} {\bibfnamefont {Zi~Hong}\ \bibnamefont
  {Liu}}, \bibinfo {author} {\bibfnamefont {Gaopei}\ \bibnamefont {Pan}},
  \bibinfo {author} {\bibfnamefont {Xiao~Yan}\ \bibnamefont {Xu}}, \bibinfo
  {author} {\bibfnamefont {Kai}\ \bibnamefont {Sun}}, \ and\ \bibinfo {author}
  {\bibfnamefont {Zi~Yang}\ \bibnamefont {Meng}},\ }\bibfield  {title}
  {\enquote {\bibinfo {title} {{Itinerant quantum critical point with fermion
  pockets and hotspots}},}\ }\href {\doibase 10.1073/pnas.1901751116}
  {\bibfield  {journal} {\bibinfo  {journal} {Proceedings of the National
  Academy of Sciences}\ }\textbf {\bibinfo {volume} {116}},\ \bibinfo {pages}
  {16760--16767} (\bibinfo {year} {2019})}\BibitemShut {NoStop}%
\bibitem [{\citenamefont {Schattner}\ \emph
  {et~al.}(2016{\natexlab{b}})\citenamefont {Schattner}, \citenamefont
  {Lederer}, \citenamefont {Kivelson},\ and\ \citenamefont
  {Berg}}]{PhysRevX.6.031028}%
  \BibitemOpen
  \bibfield  {author} {\bibinfo {author} {\bibfnamefont {Yoni}\ \bibnamefont
  {Schattner}}, \bibinfo {author} {\bibfnamefont {Samuel}\ \bibnamefont
  {Lederer}}, \bibinfo {author} {\bibfnamefont {Steven~A.}\ \bibnamefont
  {Kivelson}}, \ and\ \bibinfo {author} {\bibfnamefont {Erez}\ \bibnamefont
  {Berg}},\ }\bibfield  {title} {\enquote {\bibinfo {title} {{Ising Nematic
  Quantum Critical Point in a Metal: A Monte Carlo Study}},}\ }\href {\doibase
  10.1103/PhysRevX.6.031028} {\bibfield  {journal} {\bibinfo  {journal} {Phys.
  Rev. X}\ }\textbf {\bibinfo {volume} {6}},\ \bibinfo {pages} {031028}
  (\bibinfo {year} {2016}{\natexlab{b}})}\BibitemShut {NoStop}%
\bibitem [{\citenamefont {Lederer}\ \emph {et~al.}(2017)\citenamefont
  {Lederer}, \citenamefont {Schattner}, \citenamefont {Berg},\ and\
  \citenamefont {Kivelson}}]{Lederer4905}%
  \BibitemOpen
  \bibfield  {author} {\bibinfo {author} {\bibfnamefont {Samuel}\ \bibnamefont
  {Lederer}}, \bibinfo {author} {\bibfnamefont {Yoni}\ \bibnamefont
  {Schattner}}, \bibinfo {author} {\bibfnamefont {Erez}\ \bibnamefont {Berg}},
  \ and\ \bibinfo {author} {\bibfnamefont {Steven~A.}\ \bibnamefont
  {Kivelson}},\ }\bibfield  {title} {\enquote {\bibinfo {title}
  {{Superconductivity and non-Fermi liquid behavior near a nematic quantum
  critical point}},}\ }\href {\doibase 10.1073/pnas.1620651114} {\bibfield
  {journal} {\bibinfo  {journal} {Proceedings of the National Academy of
  Sciences}\ }\textbf {\bibinfo {volume} {114}},\ \bibinfo {pages} {4905--4910}
  (\bibinfo {year} {2017})}\BibitemShut {NoStop}%
\bibitem [{\citenamefont {Xu}\ \emph {et~al.}(2017)\citenamefont {Xu},
  \citenamefont {Sun}, \citenamefont {Schattner}, \citenamefont {Berg},\ and\
  \citenamefont {Meng}}]{PhysRevX.7.031058}%
  \BibitemOpen
  \bibfield  {author} {\bibinfo {author} {\bibfnamefont {Xiao~Yan}\
  \bibnamefont {Xu}}, \bibinfo {author} {\bibfnamefont {Kai}\ \bibnamefont
  {Sun}}, \bibinfo {author} {\bibfnamefont {Yoni}\ \bibnamefont {Schattner}},
  \bibinfo {author} {\bibfnamefont {Erez}\ \bibnamefont {Berg}}, \ and\
  \bibinfo {author} {\bibfnamefont {Zi~Yang}\ \bibnamefont {Meng}},\ }\bibfield
   {title} {\enquote {\bibinfo {title} {Non-fermi liquid at ($2+1$)$\mathrm{D}$
  ferromagnetic quantum critical point},}\ }\href {\doibase
  10.1103/PhysRevX.7.031058} {\bibfield  {journal} {\bibinfo  {journal} {Phys.
  Rev. X}\ }\textbf {\bibinfo {volume} {7}},\ \bibinfo {pages} {031058}
  (\bibinfo {year} {2017})}\BibitemShut {NoStop}%
\bibitem [{\citenamefont {Xu}\ \emph {et~al.}(2020)\citenamefont {Xu},
  \citenamefont {Klein}, \citenamefont {Sun}, \citenamefont {Chubukov},\ and\
  \citenamefont {Meng}}]{xu2020extracting}%
  \BibitemOpen
  \bibfield  {author} {\bibinfo {author} {\bibfnamefont {Xiao~Yan}\
  \bibnamefont {Xu}}, \bibinfo {author} {\bibfnamefont {Avraham}\ \bibnamefont
  {Klein}}, \bibinfo {author} {\bibfnamefont {Kai}\ \bibnamefont {Sun}},
  \bibinfo {author} {\bibfnamefont {Andrey~V.}\ \bibnamefont {Chubukov}}, \
  and\ \bibinfo {author} {\bibfnamefont {Zi~Yang}\ \bibnamefont {Meng}},\
  }\bibfield  {title} {\enquote {\bibinfo {title} {Extracting non-fermi liquid
  fermionic self-energy at $t=0$ from quantum monte carlo data},}\ }\href@noop
  {} {\  (\bibinfo {year} {2020})},\ \Eprint {http://arxiv.org/abs/2003.11573}
  {arXiv:2003.11573 [cond-mat.str-el]} \BibitemShut {NoStop}%
\bibitem [{\citenamefont {Liu}\ \emph {et~al.}(2018)\citenamefont {Liu},
  \citenamefont {Xu}, \citenamefont {Qi}, \citenamefont {Sun},\ and\
  \citenamefont {Meng}}]{PhysRevB.98.045116}%
  \BibitemOpen
  \bibfield  {author} {\bibinfo {author} {\bibfnamefont {Zi~Hong}\ \bibnamefont
  {Liu}}, \bibinfo {author} {\bibfnamefont {Xiao~Yan}\ \bibnamefont {Xu}},
  \bibinfo {author} {\bibfnamefont {Yang}\ \bibnamefont {Qi}}, \bibinfo
  {author} {\bibfnamefont {Kai}\ \bibnamefont {Sun}}, \ and\ \bibinfo {author}
  {\bibfnamefont {Zi~Yang}\ \bibnamefont {Meng}},\ }\bibfield  {title}
  {\enquote {\bibinfo {title} {Itinerant quantum critical point with
  frustration and a non-fermi liquid},}\ }\href {\doibase
  10.1103/PhysRevB.98.045116} {\bibfield  {journal} {\bibinfo  {journal} {Phys.
  Rev. B}\ }\textbf {\bibinfo {volume} {98}},\ \bibinfo {pages} {045116}
  (\bibinfo {year} {2018})}\BibitemShut {NoStop}%
\bibitem [{\citenamefont {Bauer}\ \emph {et~al.}(2020)\citenamefont {Bauer},
  \citenamefont {Schattner}, \citenamefont {Trebst},\ and\ \citenamefont
  {Berg}}]{Bauer2020}%
  \BibitemOpen
  \bibfield  {author} {\bibinfo {author} {\bibfnamefont {Carsten}\ \bibnamefont
  {Bauer}}, \bibinfo {author} {\bibfnamefont {Yoni}\ \bibnamefont {Schattner}},
  \bibinfo {author} {\bibfnamefont {Simon}\ \bibnamefont {Trebst}}, \ and\
  \bibinfo {author} {\bibfnamefont {Erez}\ \bibnamefont {Berg}},\ }\bibfield
  {title} {\enquote {\bibinfo {title} {{Hierarchy of energy scales in an O(3)
  symmetric antiferromagnetic quantum critical metal: A Monte Carlo study}},}\
  }\href {\doibase 10.1103/PhysRevResearch.2.023008} {\bibfield  {journal}
  {\bibinfo  {journal} {Phys. Rev. Research}\ }\textbf {\bibinfo {volume}
  {2}},\ \bibinfo {pages} {023008} (\bibinfo {year} {2020})}\BibitemShut
  {NoStop}%
\bibitem [{Note1()}]{Note1}%
  \BibitemOpen
  \bibinfo {note} {For the SDW model, this is technically the quasi-long ranged
  ordered phase}\BibitemShut {NoStop}%
\bibitem [{\citenamefont {Carrasquilla}\ and\ \citenamefont
  {Melko}(2017)}]{Carrasquilla2017}%
  \BibitemOpen
  \bibfield  {author} {\bibinfo {author} {\bibfnamefont {Juan}\ \bibnamefont
  {Carrasquilla}}\ and\ \bibinfo {author} {\bibfnamefont {Roger~G.}\
  \bibnamefont {Melko}},\ }\bibfield  {title} {\enquote {\bibinfo {title}
  {{Machine learning phases of matter}},}\ }\href {\doibase 10.1038/nphys4035}
  {\bibfield  {journal} {\bibinfo  {journal} {Nature Physics}\ }\textbf
  {\bibinfo {volume} {13}},\ \bibinfo {pages} {431--434} (\bibinfo {year}
  {2017})}\BibitemShut {NoStop}%
\bibitem [{\citenamefont {Carleo}\ and\ \citenamefont
  {Troyer}(2017)}]{Carleo2016}%
  \BibitemOpen
  \bibfield  {author} {\bibinfo {author} {\bibfnamefont {Giuseppe}\
  \bibnamefont {Carleo}}\ and\ \bibinfo {author} {\bibfnamefont {Matthias}\
  \bibnamefont {Troyer}},\ }\bibfield  {title} {\enquote {\bibinfo {title}
  {{Solving the quantum many-body problem with artificial neural networks}},}\
  }\href {\doibase 10.1126/science.aag2302} {\bibfield  {journal} {\bibinfo
  {journal} {Science}\ }\textbf {\bibinfo {volume} {355}},\ \bibinfo {pages}
  {602--606} (\bibinfo {year} {2017})}\BibitemShut {NoStop}%
\bibitem [{\citenamefont {Torlai}\ \emph {et~al.}(2018)\citenamefont {Torlai},
  \citenamefont {Mazzola}, \citenamefont {Carrasquilla}, \citenamefont
  {Troyer}, \citenamefont {Melko},\ and\ \citenamefont {Carleo}}]{Torlai2018}%
  \BibitemOpen
  \bibfield  {author} {\bibinfo {author} {\bibfnamefont {Giacomo}\ \bibnamefont
  {Torlai}}, \bibinfo {author} {\bibfnamefont {Guglielmo}\ \bibnamefont
  {Mazzola}}, \bibinfo {author} {\bibfnamefont {Juan}\ \bibnamefont
  {Carrasquilla}}, \bibinfo {author} {\bibfnamefont {Matthias}\ \bibnamefont
  {Troyer}}, \bibinfo {author} {\bibfnamefont {Roger}\ \bibnamefont {Melko}}, \
  and\ \bibinfo {author} {\bibfnamefont {Giuseppe}\ \bibnamefont {Carleo}},\
  }\bibfield  {title} {\enquote {\bibinfo {title} {{Neural-network quantum
  state tomography}},}\ }\href {\doibase 10.1038/s41567-018-0048-5} {\bibfield
  {journal} {\bibinfo  {journal} {Nature Physics}\ }\textbf {\bibinfo {volume}
  {14}},\ \bibinfo {pages} {447--450} (\bibinfo {year} {2018})}\BibitemShut
  {NoStop}%
\bibitem [{\citenamefont {Broecker}\ \emph {et~al.}(2017)\citenamefont
  {Broecker}, \citenamefont {Carrasquilla}, \citenamefont {Melko},\ and\
  \citenamefont {Trebst}}]{Broecker2017}%
  \BibitemOpen
  \bibfield  {author} {\bibinfo {author} {\bibfnamefont {Peter}\ \bibnamefont
  {Broecker}}, \bibinfo {author} {\bibfnamefont {Juan}\ \bibnamefont
  {Carrasquilla}}, \bibinfo {author} {\bibfnamefont {Roger~G.}\ \bibnamefont
  {Melko}}, \ and\ \bibinfo {author} {\bibfnamefont {Simon}\ \bibnamefont
  {Trebst}},\ }\bibfield  {title} {\enquote {\bibinfo {title} {{Machine
  learning quantum phases of matter beyond the fermion sign problem}},}\ }\href
  {\doibase 10.1038/s41598-017-09098-0} {\bibfield  {journal} {\bibinfo
  {journal} {Scientific Reports}\ }\textbf {\bibinfo {volume} {7}},\ \bibinfo
  {pages} {8823} (\bibinfo {year} {2017})}\BibitemShut {NoStop}%
\bibitem [{\citenamefont {Ch'ng}\ \emph {et~al.}(2017)\citenamefont {Ch'ng},
  \citenamefont {Carrasquilla}, \citenamefont {Melko},\ and\ \citenamefont
  {Khatami}}]{Chng2017}%
  \BibitemOpen
  \bibfield  {author} {\bibinfo {author} {\bibfnamefont {Kelvin}\ \bibnamefont
  {Ch'ng}}, \bibinfo {author} {\bibfnamefont {Juan}\ \bibnamefont
  {Carrasquilla}}, \bibinfo {author} {\bibfnamefont {Roger~G.}\ \bibnamefont
  {Melko}}, \ and\ \bibinfo {author} {\bibfnamefont {Ehsan}\ \bibnamefont
  {Khatami}},\ }\bibfield  {title} {\enquote {\bibinfo {title} {{Machine
  Learning Phases of Strongly Correlated Fermions}},}\ }\href {\doibase
  10.1103/PhysRevX.7.031038} {\bibfield  {journal} {\bibinfo  {journal} {Phys.
  Rev. X}\ }\textbf {\bibinfo {volume} {7}},\ \bibinfo {pages} {031038}
  (\bibinfo {year} {2017})}\BibitemShut {NoStop}%
\bibitem [{\citenamefont {Zhang}\ and\ \citenamefont {Kim}(2017)}]{Zhang2017}%
  \BibitemOpen
  \bibfield  {author} {\bibinfo {author} {\bibfnamefont {Yi}~\bibnamefont
  {Zhang}}\ and\ \bibinfo {author} {\bibfnamefont {Eun-Ah}\ \bibnamefont
  {Kim}},\ }\bibfield  {title} {\enquote {\bibinfo {title} {{Quantum Loop
  Topography for Machine Learning}},}\ }\href {\doibase
  10.1103/PhysRevLett.118.216401} {\bibfield  {journal} {\bibinfo  {journal}
  {Phys. Rev. Letters}\ }\textbf {\bibinfo {volume} {118}},\ \bibinfo {pages}
  {216401} (\bibinfo {year} {2017})}\BibitemShut {NoStop}%
\bibitem [{\citenamefont {Zhang}\ \emph {et~al.}(2019)\citenamefont {Zhang},
  \citenamefont {Bauer}, \citenamefont {Broecker}, \citenamefont {Trebst},\
  and\ \citenamefont {Kim}}]{Zhang2019}%
  \BibitemOpen
  \bibfield  {author} {\bibinfo {author} {\bibfnamefont {Yi}~\bibnamefont
  {Zhang}}, \bibinfo {author} {\bibfnamefont {Carsten}\ \bibnamefont {Bauer}},
  \bibinfo {author} {\bibfnamefont {Peter}\ \bibnamefont {Broecker}}, \bibinfo
  {author} {\bibfnamefont {Simon}\ \bibnamefont {Trebst}}, \ and\ \bibinfo
  {author} {\bibfnamefont {Eun-Ah}\ \bibnamefont {Kim}},\ }\bibfield  {title}
  {\enquote {\bibinfo {title} {{Probing transport in quantum many-fermion
  simulations via quantum loop topography}},}\ }\href {\doibase
  10.1103/PhysRevB.99.161120} {\bibfield  {journal} {\bibinfo  {journal} {Phys.
  Rev. B}\ }\textbf {\bibinfo {volume} {99}},\ \bibinfo {pages} {161120(R)}
  (\bibinfo {year} {2019})}\BibitemShut {NoStop}%
\bibitem [{\citenamefont {Wu}\ and\ \citenamefont {Zhang}(2005)}]{Wu2005}%
  \BibitemOpen
  \bibfield  {author} {\bibinfo {author} {\bibfnamefont {Congjun}\ \bibnamefont
  {Wu}}\ and\ \bibinfo {author} {\bibfnamefont {Shou~Cheng}\ \bibnamefont
  {Zhang}},\ }\bibfield  {title} {\enquote {\bibinfo {title} {{Sufficient
  condition for absence of the sign problem in the fermionic quantum Monte
  Carlo algorithm}},}\ }\href {\doibase 10.1103/PhysRevB.71.155115} {\bibfield
  {journal} {\bibinfo  {journal} {Phys. Rev. B - Condensed Matter and Materials
  Physics}\ }\textbf {\bibinfo {volume} {71}},\ \bibinfo {pages} {1--14}
  (\bibinfo {year} {2005})}\BibitemShut {NoStop}%
\bibitem [{\citenamefont {Scalapino}\ \emph {et~al.}(1993)\citenamefont
  {Scalapino}, \citenamefont {White},\ and\ \citenamefont
  {Zhang}}]{Scalapino1993}%
  \BibitemOpen
  \bibfield  {author} {\bibinfo {author} {\bibfnamefont {Douglas~J.}\
  \bibnamefont {Scalapino}}, \bibinfo {author} {\bibfnamefont {Steven~R.}\
  \bibnamefont {White}}, \ and\ \bibinfo {author} {\bibfnamefont {Shoucheng}\
  \bibnamefont {Zhang}},\ }\bibfield  {title} {\enquote {\bibinfo {title}
  {{Insulator, metal, or superconductor: The criteria}},}\ }\href {\doibase
  10.1103/PhysRevB.47.7995} {\bibfield  {journal} {\bibinfo  {journal} {Phys.
  Rev. B}\ }\textbf {\bibinfo {volume} {47}},\ \bibinfo {pages} {7995--8007}
  (\bibinfo {year} {1993})}\BibitemShut {NoStop}%
\bibitem [{\citenamefont {Binder}\ and\ \citenamefont
  {Heermann}(2010)}]{Binder}%
  \BibitemOpen
  \bibfield  {author} {\bibinfo {author} {\bibfnamefont {Kurt}\ \bibnamefont
  {Binder}}\ and\ \bibinfo {author} {\bibfnamefont {Dieter}\ \bibnamefont
  {Heermann}},\ }\href {\doibase 10.1007/978-3-642-03163-2} {\emph {\bibinfo
  {title} {Monte Carlo Simulation in Statistical Physics : An Introduction}}},\
  Vol.~\bibinfo {volume} {80}\ (\bibinfo {year} {2010})\BibitemShut {NoStop}%
\bibitem [{Note2()}]{Note2}%
  \BibitemOpen
  \bibinfo {note} {The ``cold start'' initializes the nematic pseudospins in
  one of the classical ground states of $H_b$ to bias the Monte Carlo kinetics
  and ensure that the simulations converge to the vicinity of a single local
  minimum.}\BibitemShut {Stop}%
\end{thebibliography}%
\appendix

\end{document}